\documentclass[epjST]{svjour}
\usepackage{graphicx,color}
\usepackage{amssymb,amsmath,amsfonts}
\definecolor{orange}{rgb}{1,0.5,0.0}
\begin{document}
\title{Rotational dynamics of entangled polymers}
\author{J-C. Walter\inst{1}\fnmsep\inst{2} \and  M. Laleman\inst{3}
\and M. Baiesi\inst{4}\fnmsep\inst{5} \and E. Carlon\inst{3} }
\institute{
Laboratoire Charles Coulomb UMR 5221, CNRS \& Universit\'e Montpellier 2 ,
F-34095, Montpellier, France
\and 
Laboratoire de Microbiologie et G\'en\'etique Mol\'eculaire UMR5100,
CNRS $\&$ Universit\'e Paul Sabatier Toulouse 3, F-31000 Toulouse, France
\and Institute for Theoretical Physics, KU Leuven, Celestijnenlaan 200D, 
B-3001 Leuven, Belgium
\and Department of Physics and Astronomy, University of Padua, 
Via Marzolo 8, I-35131 Padova, Italy
\and INFN, Sezione di Padova, Via Marzolo 8, I-35131 Padova, Italy
}
\abstract{
Some recent results on the rotational dynamics of polymers are reviewed
and extended. We focus here on the relaxation of a polymer, either
flexible or semiflexible, initially wrapped around a rigid rod.  We also
study the steady polymer rotation generated by a constant torque on the
rod.  The interplay of frictional and entropic forces leads to a complex
dynamical behavior characterized by non-trivial universal exponents. The
results are based on extensive simulations of polymers undergoing Rouse
dynamics and on an analytical approach using force balance and scaling
arguments. The analytical results are in general in good agreement with
the simulations, showing how a simplified approach can correctly capture
the complex dynamical behavior of rotating polymers.
} 

\maketitle

\section{Introduction}

Polymers, as every system composed by many units, may display a complex
dynamical behavior. Even single polymers can show rich and nontrivial
dynamical phases, especially if they are subject to spatial or topological
constraints~\cite{doi89,muth11}.  The impossibility to break the local
connectivity of the polymeric chains is the key to understand interesting
global rearrangements of these macromolecules.

In this paper we discuss the rotational dynamics of polymers,
where the polymers are either relaxing from a highly entangled
configuration by performing a rotational motion or forced to rotate
by an applied torque. We briefly review a series of published results
\cite{baie10a,baie13,walt13,walt14} and extend our analysis to some
new cases not considered so far.  There are two main motivations to
justify this study.  First, the rotational motion is a central instance
in polymer dynamics. Our analysis is based on the results of extensive
numerical simulations, which are supported by analytical arguments. The
comparison between the two approaches can teach us about the validity
and possible shortcomings of the approximations used in the analytical
calculations. This is a valuable input for other studies of polymer
dynamics. Second, it has applications to some interesting biological {\it
in vitro} problems: the DNA double helix melting \cite{thom93,schi14},
RNA unwinding during transcription \cite{revy04,belo14} and the closure
of denaturation bubbles in DNA~\cite{dasanna} are some examples ({\it in
vivo}, the relaxation of topological constraints of the DNA is achieved in
various ways, often involving molecular motors like chromatin remodellers,
polymerases\dots).

\begin{figure}[t]
\includegraphics[width=1.0\textwidth]{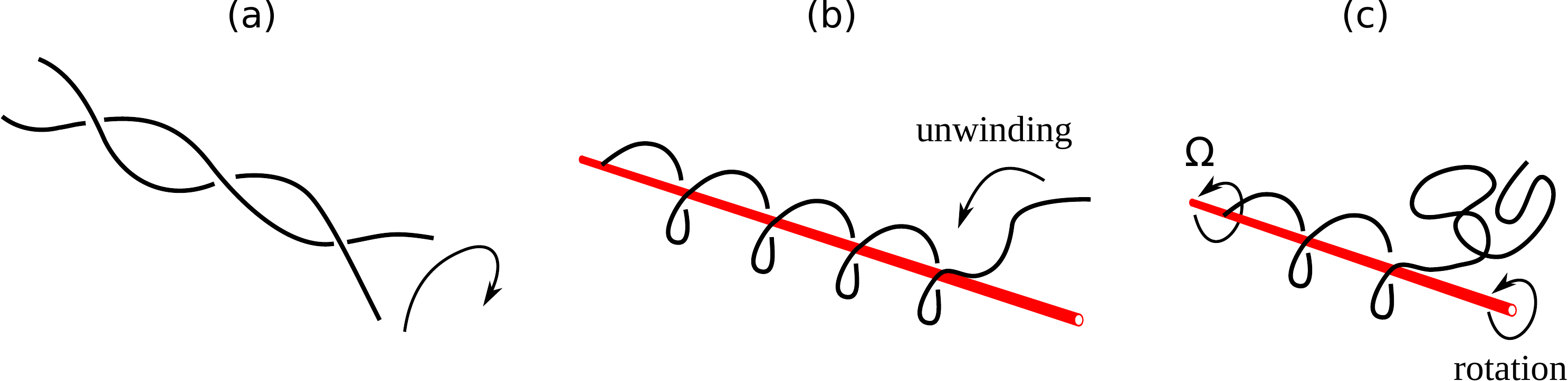}
\caption{Examples of rotational dynamics of entangled polymers.  (a)
The unwinding dynamics of two polymer chains dissociating from each
other. (b) The unwinding of a single chain from a rigid rod. (c) The
stationary conformation of a polymer attached to a rod rotating with a
constant angular velocity $\Omega$.}
\label{fig:constraints}
\end{figure}

Examples of rotational dynamics are shown in Fig.~\ref{fig:constraints}.
Consider for instance two polymers wrapped around each
other as in a double-stranded DNA helix as sketched in
Fig.~\ref{fig:constraints}(a). When brought at high temperatures or
to specific solvent conditions, DNA hydrogen bonds break~\cite{schi14}
and the two strands dissociate from each other.  The dissociation must
involve some rotational motion necessary to disentangle the two strands,
as those of Fig.~\ref{fig:constraints}(a).  Studies of DNA denaturation
dynamics have revealed that the constraint of excluded volume slows down
this relaxation~\cite{baum86,baie09,baie10a,baie13,fred14}, with time
scales that grow as a power-law of the chain length.  A simpler system is
the unwinding of a single polymer from a rigid rod~\cite{walt13,walt14},
see the sketch in Fig.~\ref{fig:constraints}(b).  Here, the advantage is
the possibility to follow the dynamics by monitoring the winding angle
of the last monomer.

The winding angle plays the role of a reaction coordinate.  Unfortunately,
this quantity cannot be defined for the unwinding of two strands. In
the case of unwinding from a rod, the knowledge of the equilibrium
statistics of the winding angle~\cite{walt11b,rudnick87} simplifies
considerably the theoretical treatment of single-polymer unwinding
dynamics~\cite{walt13,walt14,belo14}.  Here again excluded volume is
the only interaction between rod and polymer.  The unwinding dynamics
of a polymer from a rod can be viewed as a simple model of the dynamics
of a newly synthesized mRNA molecule that is clumped to the DNA it was
transcribed from~\cite{belo14}. Indeed, RNA polymerase moves as a train on
the three-dimensional railroad represented by a double stranded DNA: in
the frame where DNA is steady, the trajectory where the RNA is generated
by polymerase is a helix around the DNA backbone.  As biological polymers
are often stiff, a natural extension of the previous work is the study
of the unwinding dynamics of semi-flexible polymers, which is discussed
in this paper.

Another setup discussed in this work is shown in
Fig.~\ref{fig:constraints}(c): here the rod rotates under the influence
of a constant applied torque that leads to a constant angular velocity
$\Omega$. The polymer is attached to the surface of the rod by a single
monomer and is driven to a nonequilibrium steady rotational regime.
In this setup the impenetrable rod is fundamental to force the rotation
of the flexible polymer, as opposed to the case where a polymer with
some torsional rigidity (e.g. double stranded DNA) is set into motion
by a torque applied at one end~\cite{wolg00,wada06,wada09}.

\subsection{Polymers pulled from one end: a brief review}

The processes discussed in this paper are in a sense
the rotational counterparts of processes that have been
studied for quite some time in linearly stretched flexible
polymers~\cite{broc93,broc94,broc95,saka12,rowg12} and in semi-flexible
polymers~\cite{ober07}.

\begin{figure}[t]
\includegraphics[width=1.0\textwidth]{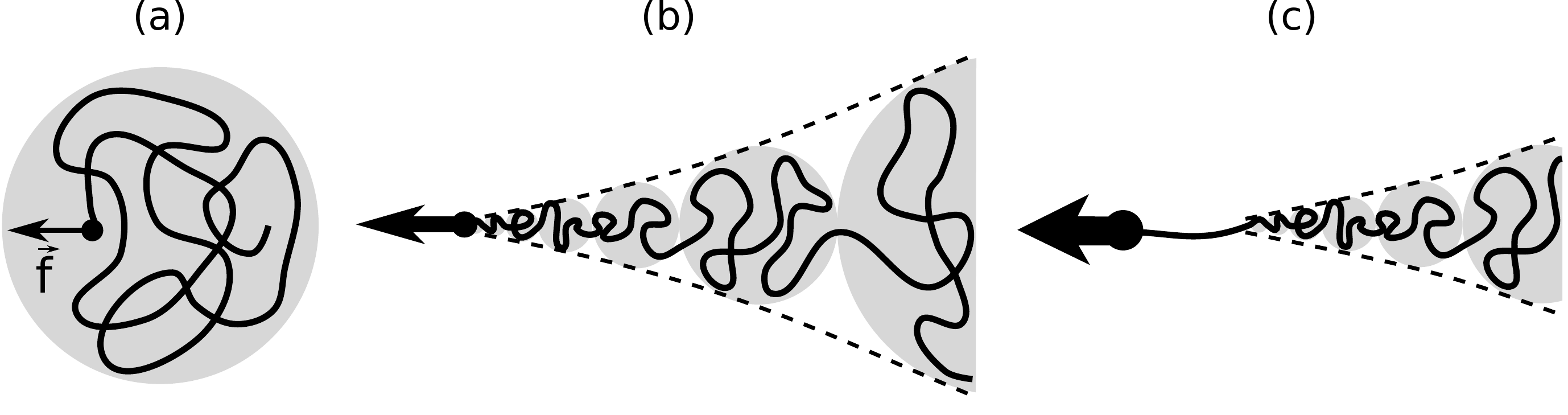}
\caption{A polymer pulled from one end by a constant force displays
three different dynamical regimes. (a) At weak forces the polymer is not
significantly perturbed with respect to the equilibrium conformation.
(b) At intermediate forces the polymer assumes the shape of a trumpet.
(c) At strong forces a part of the polymer close to the pulled end is
fully stretched, while a trumpet developes at the end part.}
\label{fig:pulled_end}
\end{figure}

Consider for instance a strong flow stretching a tethered polymer to
its full elongation. When the flow is stopped, the polymer recoils
back into its equilibrium conformation, starting from its free end:
the dynamics of Fig.~\ref{fig:constraints}(b) is the rotational
equivalent of the recoiling of a stretched polymer.  Similarly, for a
polymer forced to rotate around a rigid rod at constant angular velocity
(Fig.~\ref{fig:constraints}(c)), the counterpart is a polymer pulled from
one end by a constant force. As we will make use of scaling arguments
borrowed from the latter case, we recall briefly the three different
dynamical regimes one finds in polymers pulled from one end.  In the weak
force regime the polymer shape is not perturbed appreciably compared
to its equilibrium coiled conformation (Fig.~\ref{fig:pulled_end}(a)).
This occurs when $f \lesssim k_B T/R_F$ where $k_B$ is the Boltzmann
constant, $T$ the temperature and $R_F$ the equilibrium Flory radius
of the polymer. For a flexible polymer with $L$ monomers separated
by a distance $a$ one has $R_F = a L^\nu$, where $\nu$ is the Flory
exponent~\cite{dege79}. In an intermediate force regime, corresponding to
$k_BT/R_F \lesssim f \lesssim k_BT/a$, the polymer assumes a ``trumpet''
shape as illustrated in Fig.~\ref{fig:pulled_end}(b).  It can be
considered as being composed by a sequence of blobs of size $\xi(x)$,
which increases starting from the pulled end ($x$ is the direction along
which the force is applied).  Finally, in the strong force regime where
$f \gtrsim k_BT/a$ the pulled end is followed by a stretched portion of
polymer (Fig.~\ref{fig:pulled_end}(c)), which converts into a trumpet
shape at some point along the chain.  We will show that these concepts
enter naturally also in the characterization of rotational or unwinding
dynamics.

\subsection{Simulated models}

In order to test the universality of the results, different on- and
off-lattice models were simulated in two and three dimensions. In this
paper we focus on simulations of ideal polymers, which are more efficient
to implement. A polymer is composed of $L$ links of length $a$ that is set
to unity. Thus, in the following $L$ means the length of the polymer in
unit of $a$. In 3D the only excluded volume interactions are between the
polymer and the rod whereas in 2D excluded volume interactions are between
the polymer and a disk in the plane. In simulations, the rod is chosen
large enough to avoid the polymer to cross it during a move.  Specifically
the following models were simulated: Random Walks (RWs) on square (2D) and
on FCC (3D) lattices and the discrete worm like chain (WLC) \cite{sche74}
off-lattice in 3D.  In the WLC a polymer configuration is characterized
by angles $\alpha_i$ (with $1 <i <L$) between two consecutive bonds. Each
angle contributes to the bending energy with a term $-\cos \alpha_i$.
A configuration at inverse temperature $\beta$ thus carries a Boltzmann
weight $\exp[\beta  \sum_i \cos \alpha_i]$ where $\beta\equiv1/(k_BT)$
is the inverse temperature, $T$ the temperature and $k_B$ the Boltzmann
constant. In the following, the inverse temperature $\beta$ will be
expressed in inverse unit of $k_B$ that is set to unity. The WLC has
the property to be rigid at the scale of the persistence length $l_p$
that is an increasing function of the temperature.  At low temperature,
$l_p\sim \beta$, whereas in the high temperature limit $\beta \to 0$ one
recovers the freely jointed chain (FJC).  The dynamics was implemented
using either a kinetic Monte Carlo update, with Metropolis accepting
rules, or a Langevin thermostat~\cite{Vanden06}.  Both schemes are
implementations of Rouse dynamics~\cite{doi89}. In Monte Carlo simulations
the time unit corresponds to $L$ attempted local moves.

\begin{figure}[t]
\centering\includegraphics[angle=0,width=0.95\textwidth]{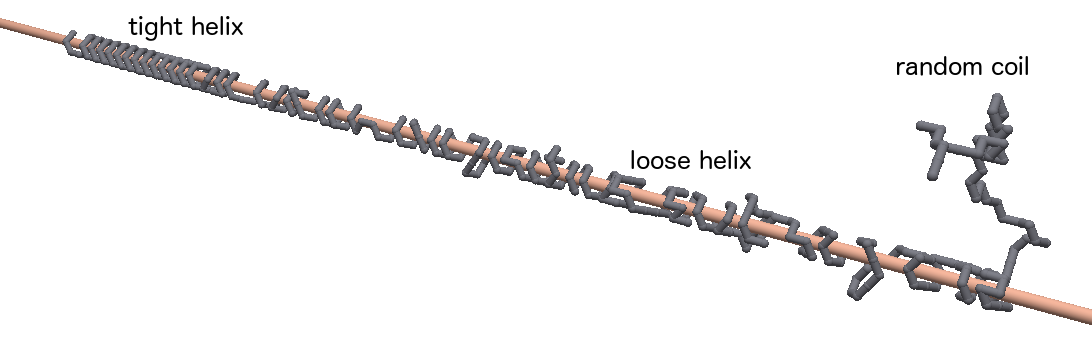}
\vskip 3mm
\centering\includegraphics[angle=0,width=0.7\textwidth]{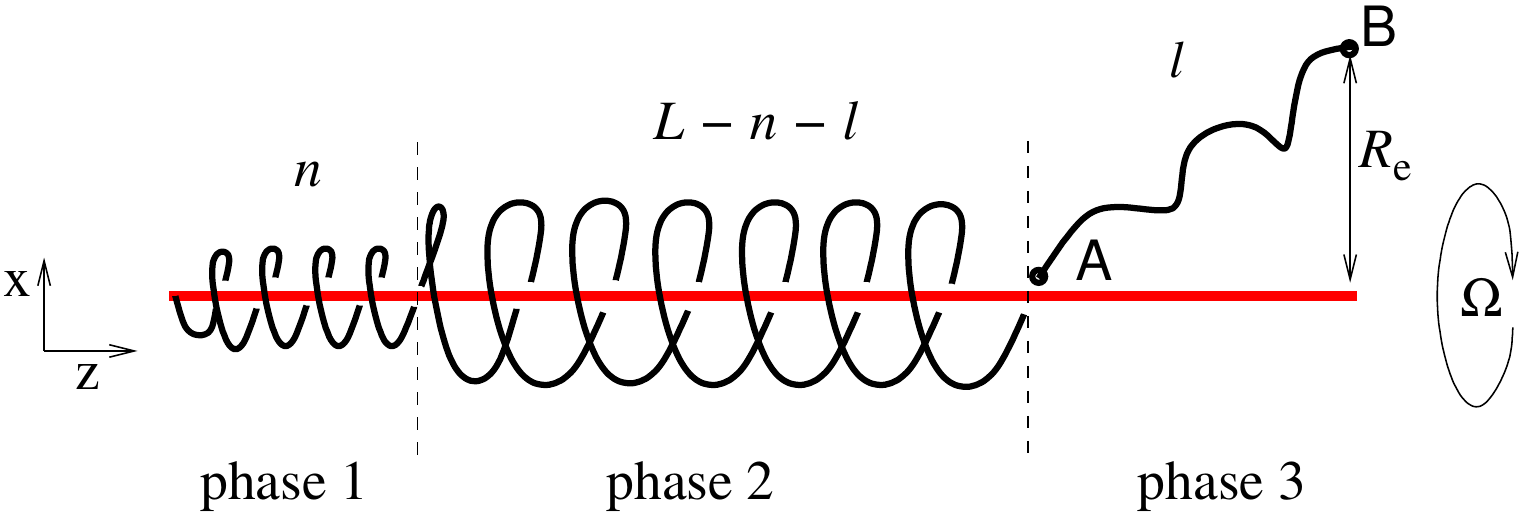}
\caption{Top: Snapshot of a simulation of a RW relaxing from an initial
conformation tightly wrapped around a rod. The snapshot suggests that
the dynamics could be modeled by a three phase model (bottom figure). The
three phase are a tight (frozen) helix, a looser helix and an end coil.}
\label{fig:rod_sketch}
\end{figure}

\section{Relaxation dynamics of a polymer wrapped around a rod}
\label{sec:relaxation_entangled}

In this section we consider the relaxation dynamics of a polymer
initially wrapped around an infinitely long rod. We start from
reviewing the predictions of the three phase model for flexible
polymers~\cite{walt13,walt14} and extend  the analysis to semiflexible
polymers.

Consider a polymer initially tightly wound around a long rod axis,
to which it is attached by one end. The free end then starts rotating
around the rod until the polymer gets fully unwound so as to maximize its
entropy. Snapshots of simulation runs \cite{walt13,walt14} suggest that
the polymer can be thought of as being composed of three different phases
as displayed in Fig.~\ref{fig:rod_sketch}: a part of the polymer close to
the end monomer attached to the rod is ``frozen" as in its initial helical
configuration (phase 1); in its central part the polymer is still wrapped
around the rod, but more loosely compared to the initial configuration
(phase 2); finally there is a terminal region detached from the rod in a
random coil configuration (phase 3). This suggests that a minimal model of
the dynamics of the system should include these three phases. To simplify
the description, we assume that phase 2 forms a homogeneous helix with a
constant pace and that the end coil assumes an equilibrium conformation
(this assumption is actually verified in simulations~\cite{walt14}). In
the course of time phase 2 and 3 grow at the cost of phase 1. However,
ultimately also phase 2 vanishes and only phase 3 remains because
random coils are the typical equilibrium configurations (with null
winding in average, but fluctuations scaling as the logarithm of the
chain length~\cite{rudnick87}).  In this paper we focus mainly on the
early stage of the unwinding, when the three phase coexist.

Two phase models have gained some popularity in recent studies of polymer
dynamics \cite{saka10,saka12,rowg12}. Such models are simple enough,
but also quite effective to capture the non-equilibrium dynamics of
polymers set into motion, for instance by a force pulling one of their
ends or during translocation.  As the dynamics involves a mechanism of
tension propagation along the polymer backbone, some time is needed before
the effect of the perturbation reaches the opposite end of the polymer.
In a two phase picture one subdivides the polymer into a moving domain
and a domain still at equilibrium, and focuses on the dynamics of the
boundary between the two phases. For instance, in a model of polymer
translocation, Sakaue \cite{saka10} identifies a moving phase close
to the pore that propagates at the expense of a rest phase located far
from the pore. Two phase models have also been used recently in studies
of polymers pulled by one end \cite{saka12,rowg12} and for DNA hairpin
dynamics \cite{fred14}. Interestingly, in our case, a minimal model of
the relaxation dynamics of a polymer wrapped around a rod requires three
phases rather than two. We are not aware of other types of three phase
models in the context of polymer dynamics.

\subsection{Theory}

We briefly review some of the predictions of the three phase model,
without entering in the details. The Appendix provides additional
schematic informations about the derivation of the main results. More
details can be found in~\cite{walt14}.  Let us consider first a fully
flexible polymer. Using a force-balance argument~\cite{walt14} one can
show that phase 1 shrinks as:
\begin{equation}
L-n \sim \sqrt{t}\,.
\label{growth1}
\end{equation}
The notation is also summarized in Fig.~\ref{fig:rod_sketch}: $L$ is
the total length of the polymer, while $n$ and $l$ are the lengths of
phase 1 and phase 3, respectively.  As phase 1 shrinks the fraction of
the polymer in phases 2 and 3 rotate in a corkscrew motion around the
rod. This is where the excluded volume of the rod-polymer system plays
a crucial role: if the polymer could trespass the rod, the initial phase
would melt immediately. In reality this does not happen and phase 2 can
increase only if phase 3 is also rotating around the rod. The unwinding of
phase 2 and 3 is thus the only mechanism that leads to a decrease of the
total winding angle.  The angular velocity is $\Omega \sim 1/\sqrt{t}$,
as obtained by differentiating Eq.~(\ref{growth1}) (see Ref.\cite{walt14}
and Appendix).  Here $\Omega$ decreases in time as during unwinding a
growing fraction of the polymer is set into motion, leading to an increase
of the friction. A scaling argument (see Ref.~\cite{walt14} and Appendix)
yields the following prediction for the growth dynamics of phase 3:
\begin{equation}
l \sim t^{1/(4\nu+2)}\,,
\label{evol_length}
\end{equation}
where $\nu$ is the Flory exponent ($\nu=1/2$ for a
Gaussian polymer and $\nu\approx 0.59$ for a self-avoiding
polymer~\cite{doi89}). Interestingly, the dynamics of the two
boundaries (between phase 1 and phase 2 and between phase 2 and phase
3) is characterized by different exponents (Eqs.~(\ref{growth1})
and (\ref{evol_length}), respectively).  Also the scaling
dynamics of the distance of the free end monomer (point $B$
in Fig.~\ref{fig:rod_sketch}) from the rod, denoted by $R_e$, can be
inferred from Eq.~(\ref{evol_length})
\begin{equation}
R_e \sim l^\nu \sim t^{1/z}\,,
\label{evol_Re}
\end{equation}
where the dynamical exponent is given by $z = 4 + 2/\nu$. 

Semiflexible polymer are characterized by an additional length scale, $l_p$,
the persistence length. At lengths $l \ll l_p$ the polymer behaves as a
stiff rod. In the initial stages of unwinding, when the phase 3 is still
short so that $l \ll l_p$, we expect the following growth law:
\begin{equation}
l \sim t^{1/6}\,.
\label{evol_length_semif}
\end{equation}
This can be obtained by formally setting $\nu=1$ in
Eq.~(\ref{evol_length}), as expected for a stiff polymer segment.
Similarly, one finds from Eq.~(\ref{evol_length_semif}):
\begin{equation}
R_e \sim l \sim t^{1/z}\,,
\label{evol_Re_semif}
\end{equation}
with $z = 6$. These formulas work as long as the phase 3 is formed by
a stiff straight segment.  We expect a crossover to the flexible case
(characterized by Eqs.(\ref{evol_length}) and (\ref{evol_Re})) when the
length of phase 3 exceeds the persistence length of the polymer, i.e.,
$l \gtrsim l_p$.

The total winding angle is defined as $\Theta \equiv 2\pi n_{\rm turns}$,
where $n_{\rm turns}$ is the number of turns (assumed to be a continuous
variable) that the polymer performs around the rod, starting from the
fixed end to the free one. It is convenient to introduce the mean density
of winding $\Delta \theta_1$ (for phase 1, the initial total winding
angle is thus $\Theta_0 = L \Delta \theta_1$) and $\Delta\theta_2$
(for phase 2). One has:
\begin{equation}
\Theta = n \Delta \theta_1 + (L-l-n) \Delta \theta_2\,,
\label{thetan2}
\end{equation}
as the coil (of length $l$) does not contribute to the winding. Combining
Eqs.~(\ref{growth1}), (\ref{evol_length}) and (\ref{thetan2}), we get:
\begin{eqnarray}
\Theta_0-\Theta &=& (L-n) (\Delta \theta_1 - \Delta \theta_2 ) 
+ l \Delta \theta_2 = A t^{1/2} + B t^{1/\alpha}\,, \nonumber\\
&=& A t^{1/2} \left( 1 + \frac{B}{A} \frac{1}{t^{1/2-1/\alpha}} \right)\,,
\label{corr_scaling}
\end{eqnarray}
where $\alpha = 4\nu+2$ ($\alpha=6$) in the flexible (semiflexible)
case, respectively. Here $A$ and $B$ are positive constants. The prediction is
that $\Theta_0 - \Theta$ scales as $\sim \sqrt{t}$, with a slowly decaying
correction term originating from the contribution of phase 3.

Note that in a fully analogous way we can define the winding angle for
any monomer. We indicate with $\theta (m)$ the local winding angle
of the $m$-th monomer, counted from the monomer fixed at the rod.
The total winding angle is then $\Theta = \theta (L)$. The three phase
model predicts the following linear piecewise profile for the local
winding angle:
\begin{eqnarray}
\theta (m) &=&
\left\{\begin{array}{ll}
 m \Delta \theta_1 &  \,\,\,{\rm if}\,\,\, m \leq n\\
 n \Delta \theta_1 + (m-n) \Delta \theta_2 &\,\,\,{\rm if}\,\,\, n \leq m \leq L-l\\
 n \Delta \theta_1 + (L-l-n) \Delta \theta_2 &\,\,\,{\rm if}\,\,\, L-l \leq m \leq L
\end{array}\right.
\label{local_wa}
\end{eqnarray}

\begin{figure}[!t]
\hbox{
\includegraphics[angle=0,width=0.525\textwidth]{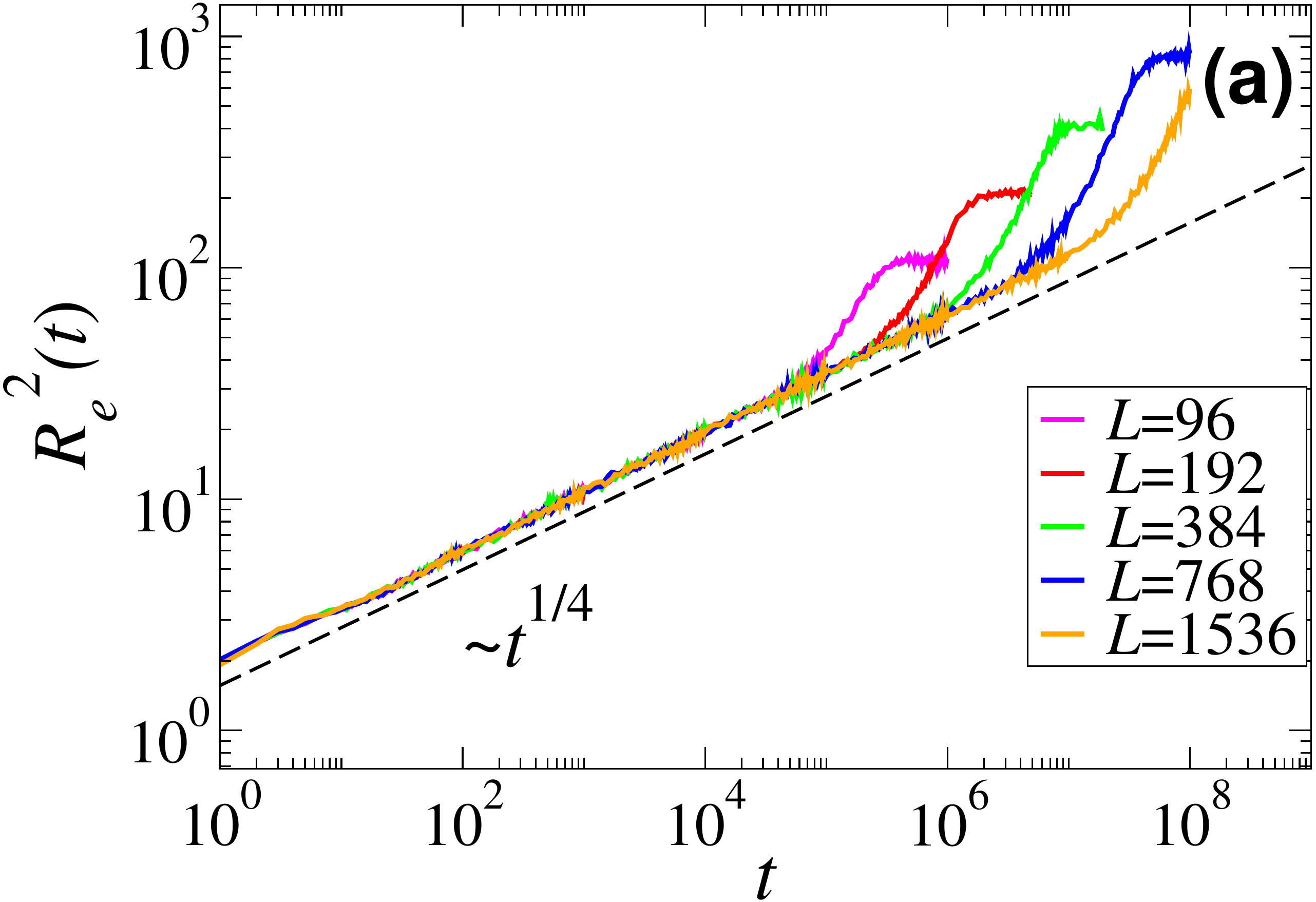}
\includegraphics[angle=0,width=0.475\textwidth]{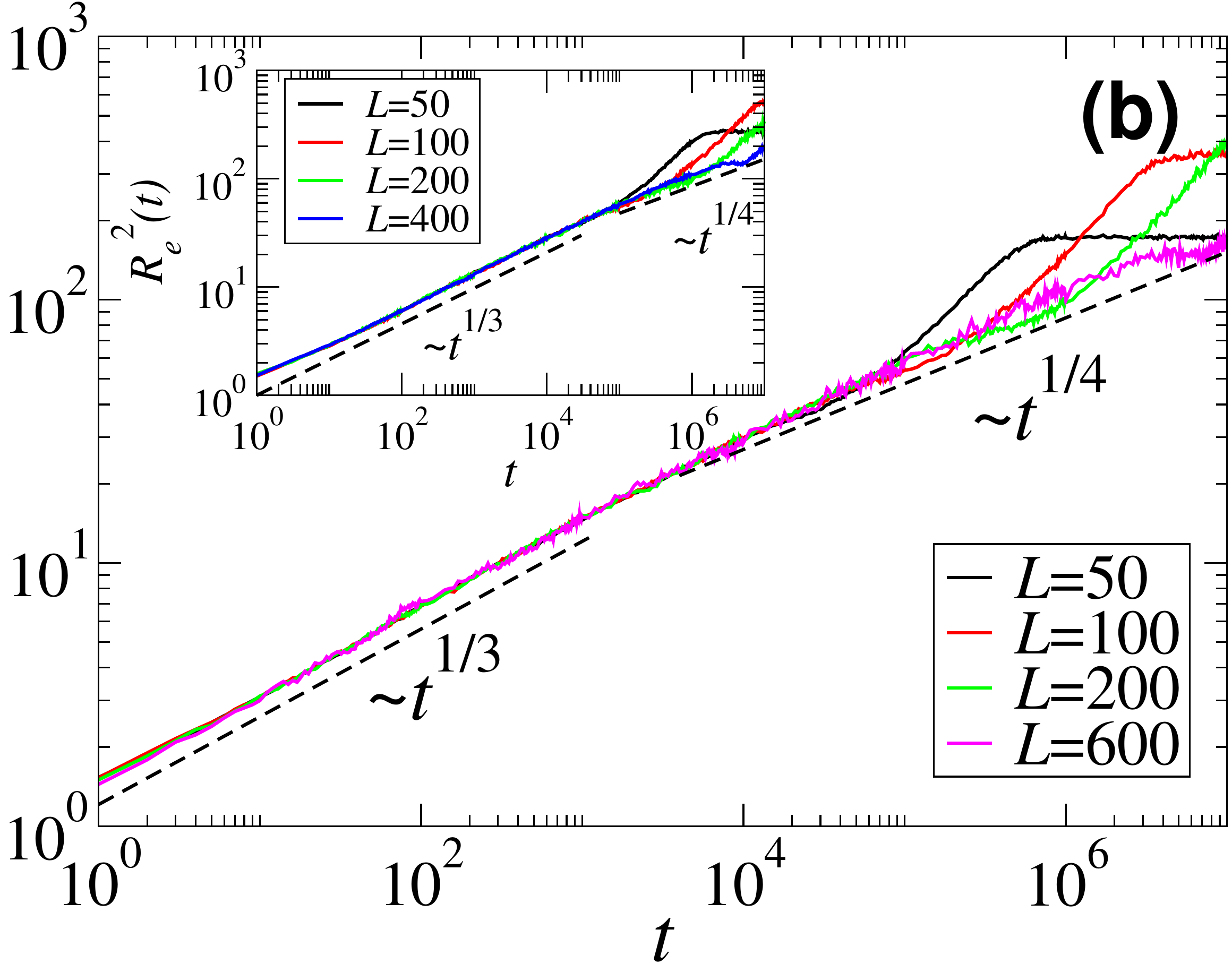}
}
\caption{
Log-log plot of the squared end-to-rod distance $R_e^2(t)$ vs.  $t$
for polymers of various lengths $L$. The panel (a) shows result for
a flexible polymer model (3D RW on a FCC lattice). The short time
regime fits the behavior predicted by Eq.~(\ref{evol_Re}) with $z=8$,
i.e. $R_e^2(t)\sim t^{1/4}$. The panel (b) displays the same quantity
for the WLC. In the main graph ($\beta=3$) it shows a crossover between
two regimes ($t\approx2\,000$).  A new regime $R_e^2(t)\sim t^{1/3}$,
i.e. $z=6$ in Eq.~(\ref{evol_Re_semif}) appears at short times. Inset: The
same quantity for the WLC at $\beta=5$. Note that at lower temperatures
the regime scaling as $R_e^2(t)\sim t^{1/3}$ extends to a wider time
interval ($t\approx50\,000$). The persistence length being larger,
the condition $l\gtrsim l_p$ occurs later, as expected.}
\label{fig:Re2}
\end{figure}

\subsection{Simulations}

Having discussed the predictions of the three-phase model we focus now
on results from simulations, which turn out to be in very good agreement
with the theory.  As we will only show results for ideal polymers, we
can set $\nu=1/2$, which is the Gaussian value. However, the results have
also been tested for self-avoiding polymers~\cite{walt14}.  We consider
first in Fig.~\ref{fig:Re2} a log-log plot of the squared end-to-rod
distance $R_e^2$ vs. time $t$ for different polymer lengths. We show
in panel (a) the flexible case with (3D RW on a FCC lattice) and (b)
the semiflexible case (WLC at $\beta=3$).  In both cases we observe
a power-law behavior at short times.  Deviations from this behavior
at longer times indicate a different type of relaxation, which is not
discussed in this article. In the flexible case (a), the dashed line
has a slope $2/z$ with $z=2/\nu+4=8$, in very good agreement with the
prediction of Eq.~(\ref{evol_Re}).  In the semiflexible case (b) one
notices a crossover between two regimes: $R_e^2 \sim t^{2/z}$ with $z=6$
at short times moving back to the flexible case $z=2/\nu+4=8$ at longer
times ($t\approx2\,000$), when the random coil becomes long enough.
In the inset of Fig.~\ref{fig:Re2}(b), the simulations are performed at
$\beta=5$ where the equilibrium persistence length $l_p$ is larger than
the main graph. As expected, we notice that the cross-over occurs at a
larger time ($t\approx50\,000$), indicating that the random coil needs
to grow larger to enter into the flexible polymer regime.

\begin{figure}[!bt]
\includegraphics[angle=0,width=0.49\textwidth]{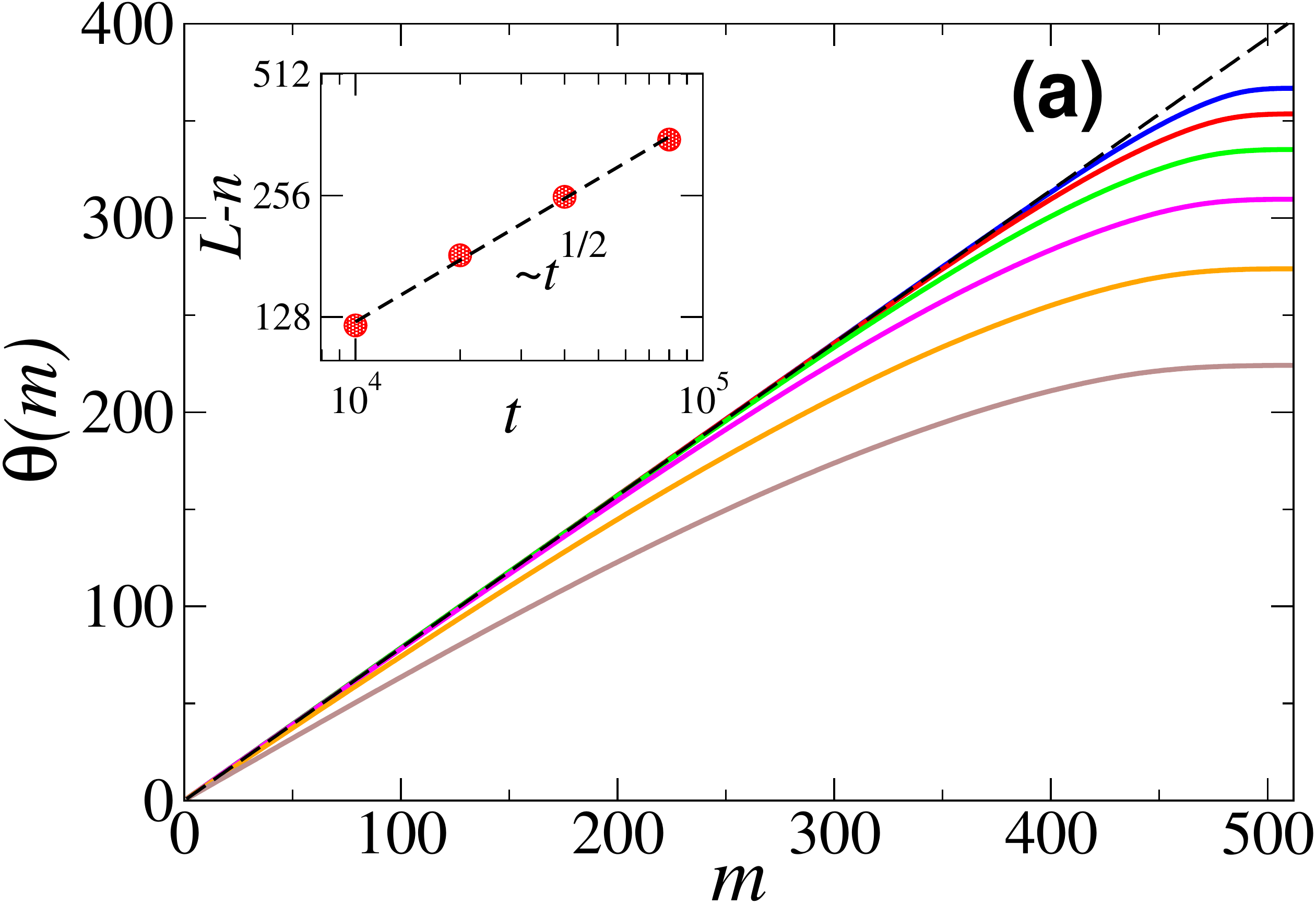}
\includegraphics[angle=0,width=0.48\textwidth]{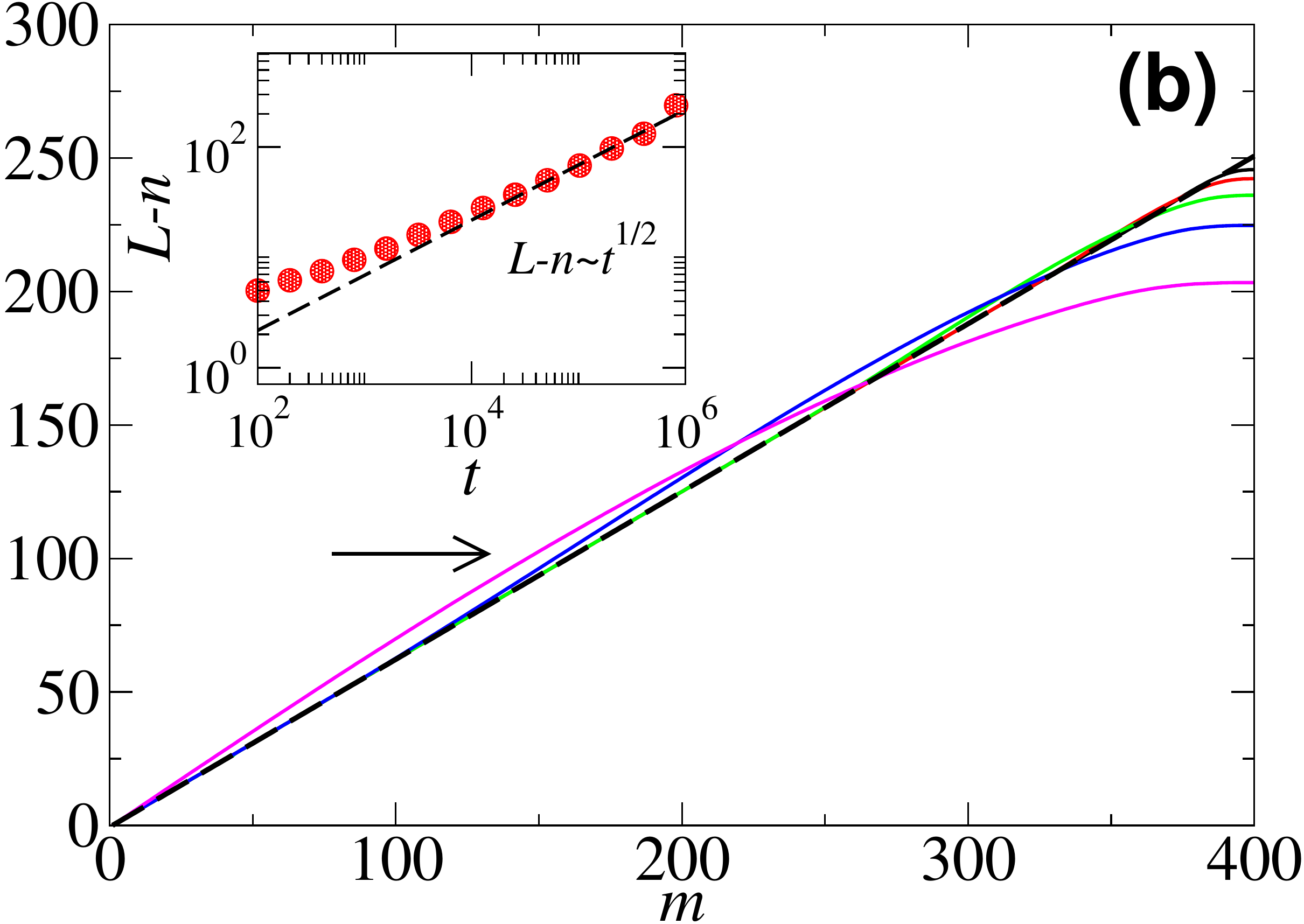}
\caption{Plot of local winding angle $\theta(m)$ vs. monomer number $m$
for different times for (a) the flexible case (2D RW of length $L=512$
on a square lattice, curves for $t=10^4$ up to $t=32\times 10^4$ doubling
$t$) and (b) the WLC at $\beta=2$ and for a length $L=400$. From top
to bottom, the times run from $t=100\times2^{4}$ to $100\times2^{12}$,
multiplying $t$ by 4. The horizontal arrows point (for the three
largest times only) to the start of the slight increase of $\theta(m)$
in the phase 1, which does not occur in (a).  (Insets) Log-log plot
of the length $L-n$ of phase 2, vs. $t$ for the (a) flexible and (b)
semiflexible case. The dashed lines have slope $1/2$, as predicted from
Eq.~(\ref{growth1}).}
\label{fig:local_winding}
\end{figure}

We continue the analysis of simulation results to test of the validity
of Eq.~(\ref{growth1}). To identify the boundary between phase 1 and
phase 2, we consider the local winding angle $\theta(m)$ as a function
of the monomer number $m$.  A plot of $\theta(m)$ vs. $m$ for different
times is shown in Fig.~\ref{fig:local_winding} for (a) a flexible
polymer (2D RW of length $L=512$) and for (b) a semiflexible polymer
(WLC of length $L=400$). At $t=0$ the polymer is prepared in phase 1
so the local winding angle is $\theta(m) = m \Delta \theta_1$ (dashed
lines in both figures).  At intermediate times the three phase model
predicts that the local winding angle is a piecewise linear function
(Eq.~(\ref{local_wa})).  The numerical results for $\theta(m)$ instead
do not show sharp boundaries: the average over different simulation runs
produces a smooth continuous curve.  We note that while in the flexible
case the local winding angle decreases monotonically in time, this is
not the case for the semiflexible polymer. Starting from the $t=0$ tight
helix configuration (dashed line in Fig.~\ref{fig:local_winding}(b))
there is a slight increase in $\theta(m)$ in the phase 1, as indicated
by the horizontal arrows (three largest times only).  This appears to be
a small effect though. To determine $n(t)$, the length of the phase 1,
we took in the semiflexible case the intersection of the $\theta(m)$
vs. $m$ curve with the line $\theta(m) = m \Delta \theta_1$.  In the
flexible case we took the value of $m$ from which $\theta(m)$ drops to
$95\%$ of the initial $t=0$ value.  In both cases the data follow the
expected $\sqrt{t}$ behavior (see insets of Fig.~\ref{fig:local_winding}).
The results are in agreement with Eq.~(\ref{growth1}).

\begin{figure*}[t]
\hbox{
\includegraphics[angle=0,height=4.5cm]{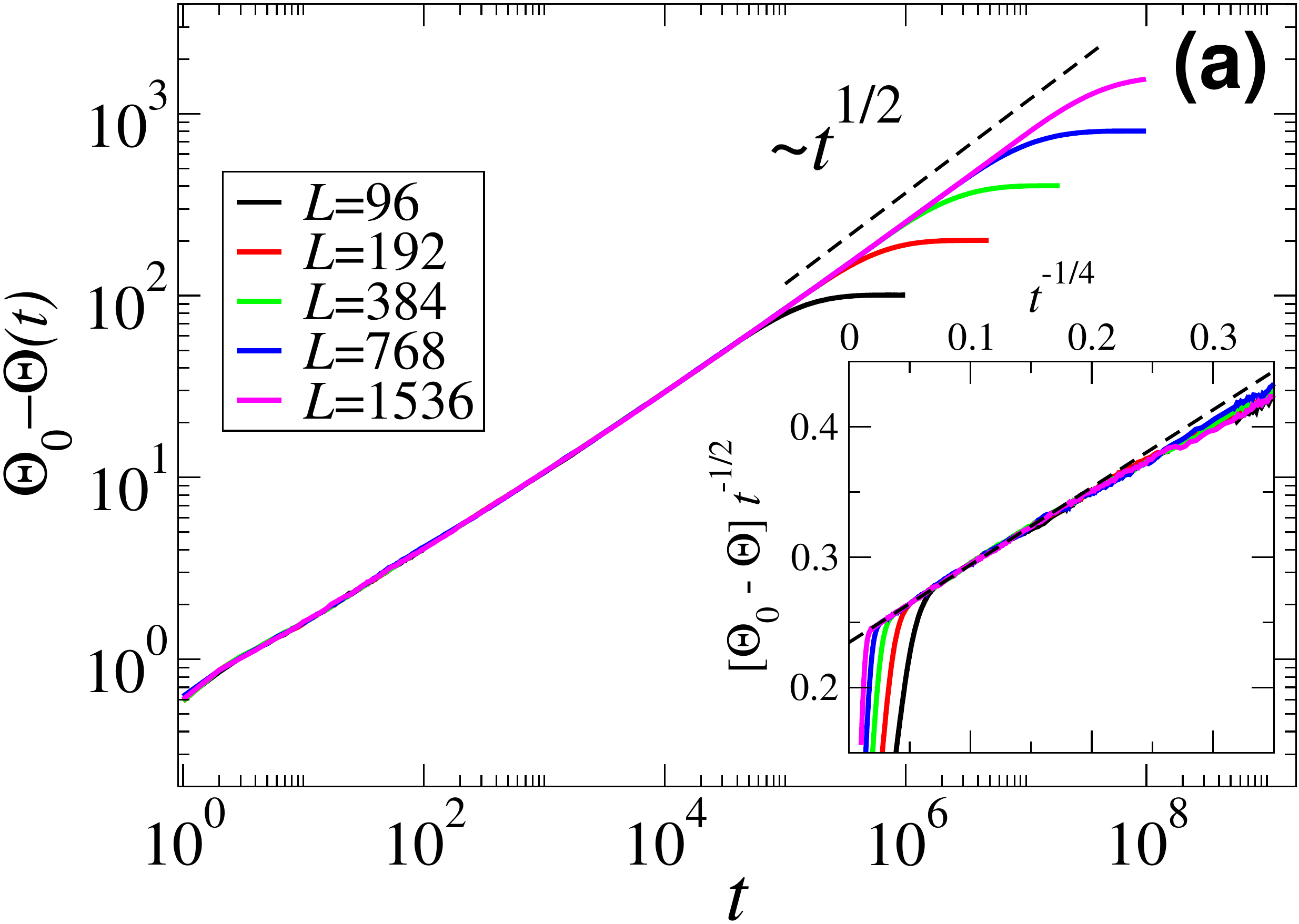}
\includegraphics[angle=0,height=4.5cm]{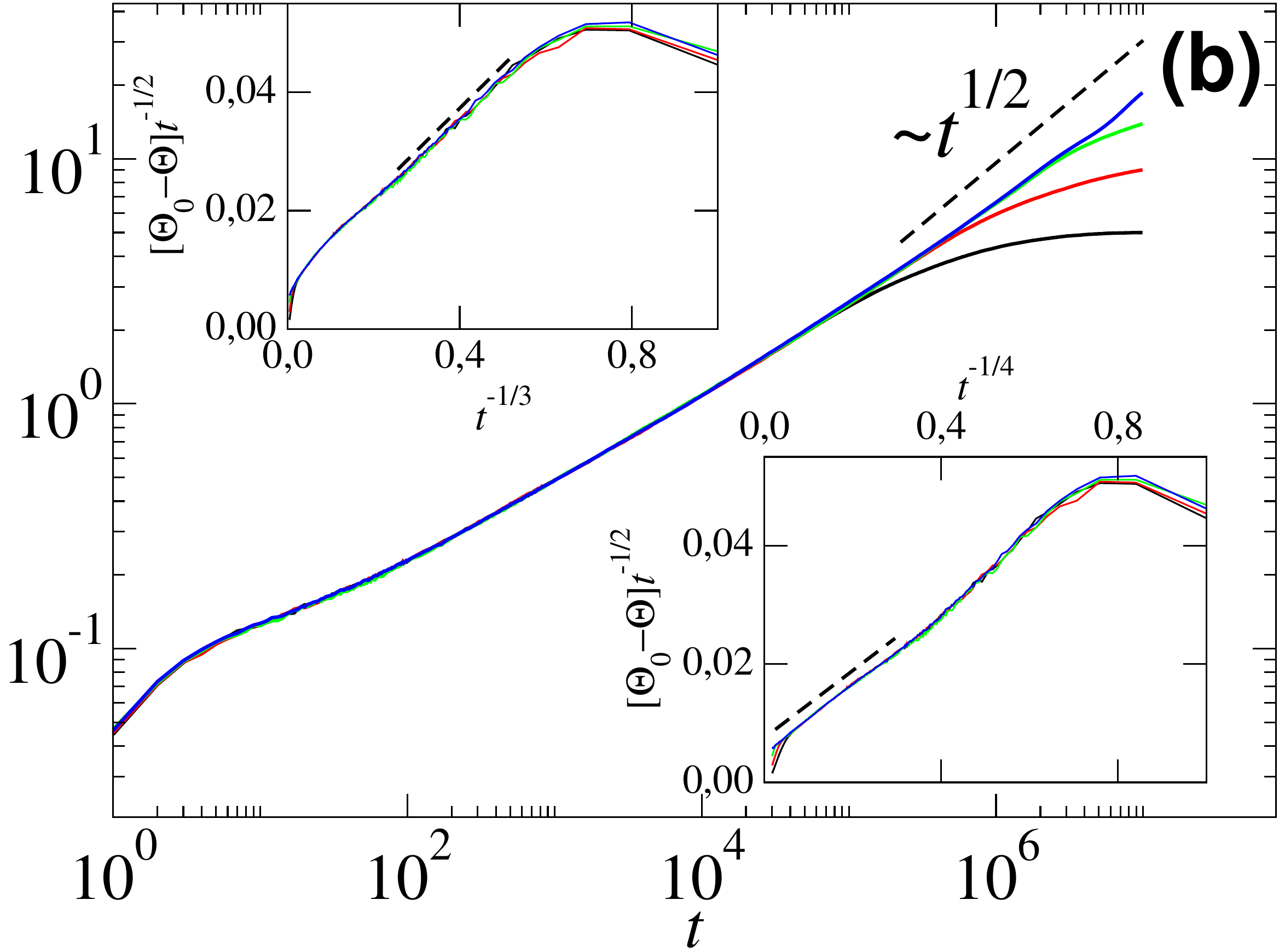}
}
\caption{Amount of unwinding for (a) a flexible polymer (RW on a FCC
lattice) and (b) a semiflexible polymer (WLC at $\beta=5$).  Insets:
corrections to scaling.  In (b) the two insets show that there is a
crossover also in the correction to scaling from the exponent $-1/3$
at small time and $-1/4$ at large time.}
\label{fig:Theta}
\end{figure*}

We focus next on the dynamics of the total winding angle
and test the validity of Eq.~(\ref{corr_scaling}). A log-log
plot of $\Theta_0 - \Theta$ as a function of time is shown in
Fig.~\ref{fig:Theta}. This quantity approaches the $\sqrt{t}$ behavior
predicted by Eq.~(\ref{corr_scaling}), but with strong corrections
to the dominant scaling. To investigate the leading corrections to
scaling reported in Eq.~(\ref{corr_scaling}), we plot in the insets
$(\Theta_0 - \Theta)/\sqrt{t}$ as a function of $1/t^{1/2-1/\alpha}$.
According to Eq.~(\ref{corr_scaling}), this quantity should follow a
straight line. This is indeed the case as can be seen in the inset of
Fig.~\ref{fig:Theta}(a). Note that for the flexible case the correction
term is $t^{-1/4}$ (for RW, $\nu=1/2$) while it is equal to $t^{-1/3}$
for semiflexible polymers.  In the insets of Fig.~\ref{fig:Theta}(b),
we observe the crossover between the term $t^{-1/3}$ at small time and
$t^{-1/4}$ at large time, confirming the prediction of our model.

\section{Stationary rotating polymer}

We focus now on the case of a polymer fixed by one end monomer
to a rod to which a constant torque is applied, as sketched in
Fig.~\ref{fig:constraints}(c).  We restrict ourselves to the
flexible case. After a transient regime, the polymer enters a
stationary nonequilibrium state where it rotates around the rod with
a constant angular velocity $\Omega$. Analogous to the case of a
polymer pulled by a constant force by one end, we expect that a weak
torque would not be able to perturb the polymer significantly from its
equilibrium conformation. The relation (\ref{final}), derived in the
Appendix, characterizes the weak torque regime. This inequality can
be interpreted as follows: polymers rotating with an angular velocity
$\Omega$, but shorter than a critical length 
\begin{equation}
L_c = \left(\frac{k_B T }{ a \gamma_0 \Omega}\right)^{1/(1+2\nu)}, 
\label{def_Lc}
\end{equation}
will rotate around the rod maintaining their random
coil shape. Polymers with length $L > L_c$ will instead be partially
wrapped around the rod: only the $L_c$ monomers at the free end may form
an equilibrated coil. To test the validity of these scaling arguments
we analyzed the behavior of the squared end distance from the rod,
$R^2_e$, as a function of $\Omega$. In the weak torque regime the full
chain rotates in an equilibrium configuration around the rod, and thus
$R_e^2 \sim L^{2\nu}$. In the high torque regime only $L_c$ monomers
can be in an equilibrium configuration, hence $R_e^2 \sim L_c^{2\nu}
\sim \Omega^{-2\nu/(1+2\nu)}$. The two regimes can be connected by means
of a scaling function:
\begin{equation}
R_e^2 = L^{2\nu}\, g\left(L  \Omega^{1/(1+2\nu)}\right)
\label{Re2_scaling}
\end{equation}
where the natural scaling variable is $x=L \Omega^{1/(1+2\nu)}$.
The scaling function in the large $\Omega$ limit behaves as $\lim_{x
\to \infty} g(x) = x^{-2\nu}$, while it tends to a constant limit for
small values of its argument.

\begin{figure}[t]
\begin{center}
\includegraphics[height=4.5cm]{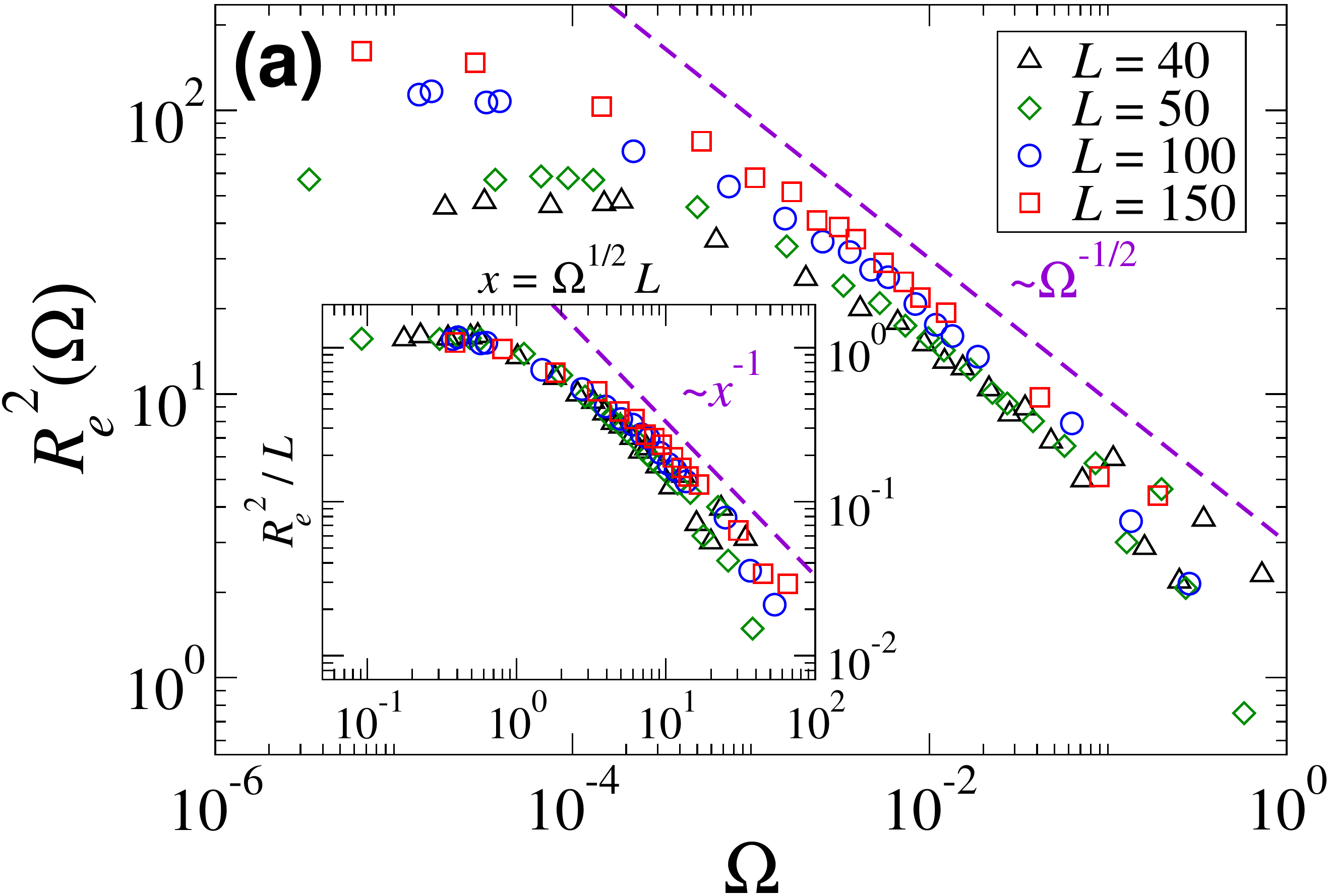}
\includegraphics[height=4.5cm]{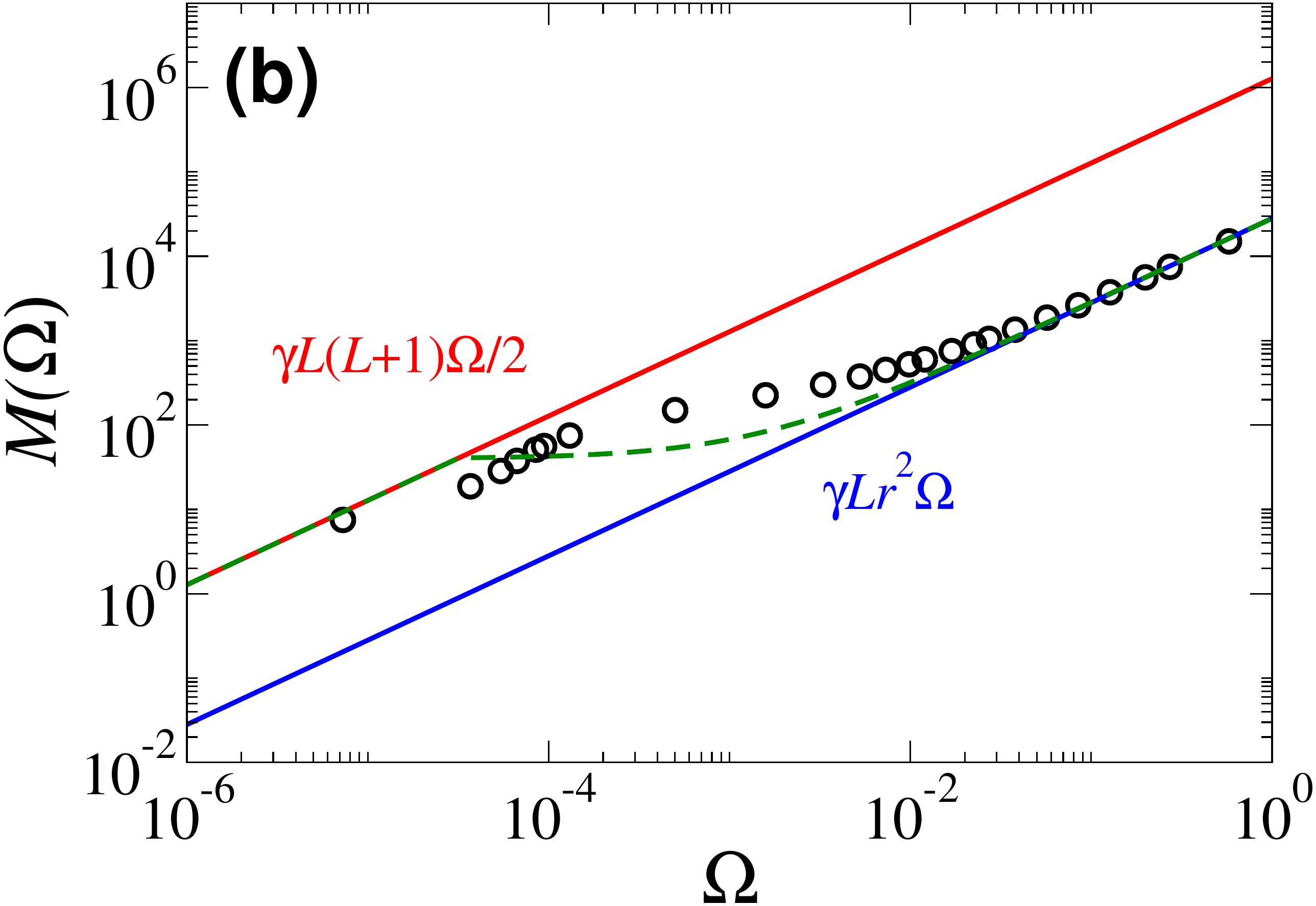}
\caption{(a) The end-to-rod distance as a function of the angular velocity
for differents sizes. For low angular velocities the polymer rotates in
equilibrium configuration around the rod. For higher velocities the two
phase model predicts that $R_{e}$ should scale as $\Omega^{-0.5}$. Inset:
rescaling according to Eq.~(\ref{Re2_scaling}). (b) Torque vs.~angular
velocity, for $L=50$. Only for low and high angular velocities a
linear relation is expected, at intermediate values of $\Omega$ a more
complicated behavior emerges.  The straight lines represent predicted
asymptotic scalings, while the dashed line is the scaling from a simple
two phase model.
}
\label{fig:R2ee_ifo_omega}
\end{center}
\end{figure}

We simulate a polymer composed by $L$ beads on a continuum using
a Langevin thermostat~\cite{Vanden06}. Each bead interacts with its
neighbors along the polymer via a FENE potential. The interaction with
the rod is modeled by a repulsive truncated Lennard-Jones potential
with radius $r_{\rm rod}=0.75a$. The distance $a$ between beads
was normalized to one, just as $k_B$ and the mass of a bead $m$. The
equations of motion were integrated using a time step of $\Delta t=0.001$.
The bath was at a temperature $T=80$ and the friction coefficient set
to $\gamma_0=10^3$. The FENE potential had a spring constant of $2.36
\cdot 10^{5}$ and a maximal extensibility of $1.5a$.  In the simulation
we apply a constant force $F$ perpendicular to the axis of the rod to
one end monomer. This monomer is constrained to move on a radius $r_{\rm
rod}$ from the axis of the rod. The applied torque $M=F r_{\rm rod}$
induces a constant angular velocity to the polymer which can be obtained
by averaging over the velocities of all monomers, once the polymer
has reached the stationary regime. Figure~\ref{fig:R2ee_ifo_omega}(a)
shows a log-log plot of the squared end-to-rod distance as a function of
$\Omega$ for polymers of various lengths. The dashed line, in agreement
with the data at high $\Omega$, corresponds to a scaling $R^2_e \sim
1/\sqrt{\Omega}$ which is the behavior predicted from the scaling
argument derived above (again we use $\nu=1/2$, which is the value for
a gaussian polymer).  In addition appropriately rescaled data collapse
in good agreement with the scaling form (\ref{Re2_scaling}), as shown
in the inset of Fig.~\ref{fig:R2ee_ifo_omega}(a).

A rigid body rotating in a viscous medium under the effect of a torque $M$
is characterized by a linear relationship between torque and angular velocity
$\Omega$
\begin{equation}
M = \Gamma \Omega,
\end{equation}
where $\Gamma$ is the rotational friction. We can estimate $\Gamma$
for a polymer rotating around the rod for some specific polymer
conformations. Let us suppose first that the polymer is in an
equilibrated coiled conformation, which is what we expect to observe at
low $\Omega$'s. In equilibrium the $l$-th monomer is at an average distance 
$r^2 (l) = a^2 l$ from the rod. Assuming that each monomer contributes a term
$\gamma_0 r^2 (l)$ to the friction we get
\begin{equation}
\Gamma_{\rm coil} = \sum_{l=1}^L \gamma_0 a^2 l = \frac{\gamma_0 a^2}{2}L(L+1).
\label{gamma_coil}
\end{equation}
As a second case consider a polymer fully and tightly wrapped around the
rod, i.e. as in a helix where each monomer is at a distance
$r_{\rm rod}$ from the rod. The rotational friction is equal to
\begin{equation}
\Gamma_{\rm helix} = \gamma_0 r_{\rm rod}^2 L\,.
\label{gamma_helix}
\end{equation}
In Fig.~\ref{fig:R2ee_ifo_omega}(b) we show a log-log plot of $M$ vs.
$\Omega$ as obtained from simulations.  The two solid lines are $M =
\Gamma_{\rm coil} \Omega$ and $M = \Gamma_{\rm helix} \Omega$ from
Eqs.~(\ref{gamma_coil}) and (\ref{gamma_helix}) without any fitting
parameters. The simulation data interpolate between these two limits: 
as expected the polymer is coiled at low torques and gets fully wrapped
around the rod at high torques.
We can get an interpolating formula for $\Gamma$ in the intermediate
regime using a two phase model. We assume that there are two distinct
contributions to the friction coming from an end coil of length $l$
and of a wrapped part of length $L-l$ leading to:
\begin{equation}
\Gamma_{\rm tot}(L,l) = \Gamma_{\rm coil}(l) + \Gamma_{\rm helix}(L-l).  
\label{eq:two_Gamma}
\end{equation}
We use Eq.~(\ref{def_Lc}) to estimate $l$ and set $l=\tilde{L}_c
=\min (L_c,L) $. In Fig.~\ref{fig:R2ee_ifo_omega}(b) we plot the curve
$M=\Gamma_{\rm tot}(L,\tilde{L}_c) \Omega$, which is shown as a dashed
line. Again, there are no fitting parameters in this computation. Note
that $\Gamma_{\rm tot}$ depends on $\Omega$ through $L_c$, see
Eq.~(\ref{def_Lc}). The analytical calculation reproduces qualitatively
the crossover behavior between the two regimes, although there are
some systematic deviations. In particular, the analytical computation
underestimates the friction.This is because the helical phase is assumed
to be tightly wrapped around the rod with all monomers at distance
$r(l)=r_{\rm rod}$ from the rod axis. In reality, the monomers are
expected to be more loosely wrapped around the rod, producing a larger
friction than the estimate of Eq.~(\ref{gamma_helix}).
Future work will focus on a construction of a better approximation 
for $\Gamma$.

\section{Conclusions}

In this paper we have reviewed some recent results about the rotational
dynamics of a single flexible polymer around a rod~\cite{walt13,walt14}
and extended the analysis to new cases. Two situations have been
discussed: the relaxational dynamics of a polymer initially wrapped around
a rigid rod and a polymer forced to rotate by a constant applied force. In
the former case we have shown how a force-balance argument, relying on a
three phase model, allows to derive the universal exponents describing the
relaxational dynamics of various quantities as the end-distance from the
rod or the winding angle. The results have been extended here to the case
of semiflexible polymers which show a crossover between different regimes.
In general, a very good agreement between theory and simulations is found.
In a polymer rotating under the influence of an applied torque we have
analyzed the dependence of the end-distance from the rod on the angular
velocity, which matches the predictions from scaling theory.  We have
shown that there is a non-linear relationship between torque and angular
velocity which is due to a conformational transition in the polymer from
a coiled to a wrapped state. A two phase model calculation, without any
adjustable parameters, reproduces semi-quantitatively the torque vs.
angular velocity curves obtained from simulations.

In conclusion, although the models discussed are rather simple, our
analysis shows that there is an underlying rich and complex dynamics.
The analytical and scaling arguments developed to study these simple
systems can guide us towards the understanding of more complex cases
of polymer dynamics. The study of polymers disentangling from linear
objects is also of interest for a better understanding of some aspects
of RNA and DNA dynamics.

\section*{Acknowledgements}

We thank G. Barkema and H. Schiessel for collaborations on the unwinding
relaxationd dynamics of flexible polymers.  J-CW is supported by the
Laboratory of Excellence Initiative (Labex) NUMEV, OD by the Scientific
Council of the University of Montpellier 2.  MB and J-CW acknowledge
the kind hospitality of the Institute of Theoretical Physics at the KU
Leuven, where part of this work was done.

\section*{Appendix}

We provide some details about the derivation of the results of
Section~\ref{sec:relaxation_entangled} (see also Ref.~\cite{walt14}).
To describe the dynamics of the polymer rotating around the rod we
neglect first the contribution of phase 3. The total free energy of the
system is given by ${\cal F}(n) = f_1 n + f_2 (L-n)$, where $f_1$ and
$f_2$ are the free energies per unit of polymer length for the phase 1
and 2 respectively. When phase 1 shrinks the phases 2 and 3 rotate in
a corkscrew motion; neglecting phase 3, the friction associated with
the dynamics comes from phase 2 and scales as the length of this phase
$\gamma_2 \sim L-n$.  The balance of frictional and entropic forces gives:
\begin{equation}
\gamma_2 \frac{dn}{dt} = - \frac{\partial {\cal F}}{\partial n}
\end{equation}
Using the above forms for $\gamma_2$ and ${\cal F}$ one can easily
integrate the previous differential equation to get the square root growth
predicted by Eq.~(\ref{growth1}).  To derive Eq.~(\ref{evol_length})
we use the analogy with a polymer pulled from one end by a constant
force $f$. As discussed in the introduction the weak force regime of
Fig.~\ref{fig:pulled_end}(a) is given by
\begin{equation}
f R_F \lesssim k_B T
\label{eq:RF}
\end{equation}
Consider a polymer rotating around an axis with an angular
velocity $\Omega$. The average distance from the axis is $R_F$,
so the linear velocity is $v \approx \Omega R_F$ and the frictional
force is thus $f \approx \gamma \Omega R_F$. In Rouse dynamics the
friction is proportional to the polymer length $\gamma \approx \gamma_0
l$~\cite{doi89}. Summarizing, the inequality (\ref{eq:RF}) for a rotating
polymer becomes:
\begin{equation}
a \gamma_0 \, l^{1+2\nu} \Omega \lesssim k_B T .
\label{final}
\end{equation}
Hence a polymer of length $l$ rotating around its axis with an angular
velocity $\Omega$ sufficiently small such that Eq.~(\ref{final})
is satisfied will maintain its equilibrium shape. If $\Omega$ is
large enough such that Eq.~(\ref{final}) is not fulfilled, then
part of the polymer will be wrapped around the rod.  For a polymer
relaxing from an entangled state around a rod, we
need the time evolution of $\Omega$ stated in 
Section~\ref{sec:relaxation_entangled}. 
Using Eq.~(\ref{thetan2}), and assuming a faster growth 
for $L-n$ compared to $l$, we find to leading order in $t$:
\begin{equation}
\Omega=\frac{d\Theta}{dt}\sim \frac{dn}{dt}\sim\frac{1}{\sqrt t}\,,
\end{equation}
where we have used Eq.~(\ref{growth1}). Then, one can use the
relation (\ref{final}) in the form of an equality to obtain
Eq.~(\ref{evol_length}).

\bigskip


\end{document}